\documentclass[twocolumn,pra,aps, twocolumn]{revtex4}

\usepackage[utf8]{inputenc}
\usepackage[english]{babel}
\usepackage[T1]{fontenc}
\usepackage{hyperref}
\usepackage{braket}

\usepackage{amsmath,amsfonts,amsthm,amssymb,amsbsy}
\usepackage{graphicx}
\usepackage{amssymb}
\usepackage{color}
\usepackage{hyperref}

\newtheorem{theorem}{Theorem}
\newtheorem{lemma}{Lemma}
\newtheorem{conjecture}{Conjecture}

\newcommand{\hc}{\mathrm{h.c.}}

\newcommand{\vac}{\ket{\mathrm{vac}}{}}
\newcommand{\uu}{\mathcal{\hat U}}

\newcommand{\wo}{x}
\newcommand{\wt}{y}

\newcommand{\eq}[1]{\begin{align}#1\end{align}}
\newcommand{\seq}[1]{\begin{subequations}#1\end{subequations}}
\newcommand{\sseq}[1]{\seq{\eq{#1}}}

\begin{document}

\title{Gaussian functions are optimal for waveguided nonlinear-quantum-optical processes}

\author{Nicol\'as Quesada }

\affiliation{Perimeter Institute for Theoretical Physics, Waterloo, Ontario, N2L 2Y5, Canada}
\author{Agata M. Bra\'nczyk}
\email[]{abranczyk@perimeterinstitute.ca }
\affiliation{Perimeter Institute for Theoretical Physics, Waterloo, Ontario, N2L 2Y5, Canada}

\begin{abstract}

Many nonlinear optical technologies require the two-mode spectral amplitude function that describes them---the \emph{joint spectral amplitude} (JSA)---to be separable. We prove that  the JSA factorizes \emph{only} when the incident pump field and phase-matching function are Gaussian functions.  We show this by mapping our problem to a known result, in continuous variable quantum information, that only  squeezed states remain unentangled when combined on a beam splitter. We then conjecture that only a squeezed state minimizes entanglement when sent through a  beam splitter with another pre-specified ket. This implies that to maximize JSA separability when one of the (pump or nonlinear medium) functions is non-Gaussian, the other function \emph{must} be Gaussian. This answers an outstanding question about optimal design of certain nonlinear processes, and is of practical interest to researchers using waveguide nonlinear optics  to generate and manipulate quantum light. 
\end{abstract}

\maketitle

\section{Introduction}

Nonclassical light is used in many emerging optical technologies. It is the ideal carrier of quantum information, either over long distances for quantum  communication \cite{Gisin2007} or quantum cryptography \cite{Bennett1984}, or between neighbouring modules within quantum information processing devices \cite{kok2010introduction}. It can  enhance metrology \cite{Higgins2007,Nagata2007},  imaging \cite{Brida2010},  lithography \cite{Boto2000}, and optical coherence tomography \cite{Nasr2008}. Progress in these technologies requires  high-quality generation and manipulation of quantum-light-sources.

Many physical systems have been explored as sources of non-classical light. Single-photon emitters, such as trapped single neutral atoms \cite{Beugnon2006}, ions \cite{Keller2004}, atomic ensembles \cite{Kuzmich2003}, quantum dots \cite{Santori2002,Reimer2012}, and nitrogen vacancy centres in diamond \cite{Kurtsiefer2000}, can be made to generate photons ``on-demand'', and are thus very promising. But they are challenging to work with as they require expensive and bulky refrigeration to reduce thermal noise. Nonlinear optical sources \cite{christ2013single,Cohen2009}, on the other hand, are relatively inexpensive, easy to use, and robust compared to single-photon emitters. They generate photons spontaneously, but can be made near-deterministic through multiplexing \cite{Pittman2002,Jeffrey2004}. They are more versatile for shaping the photon's spectro-temporal properties, and can be used to generate other non-classical states of light such as entangled photon pairs \cite{kwiat1995new,Kwiat1999a}, squeezed states \cite{Lvovsky2014}, thermal states \cite{Lvovsky2014}, and cat states \cite{Ourjoumtsev2006}. Furthermore, nonlinear sources can transform existing non-classical light \cite{ansari2017temporal,eckstein2011quantum,brecht2014demonstration}.  

We focus on the class of second-order, i.e. $\chi^{(2)}$, nonlinear processes. The most popular $\chi^{(2)}$ process, spontaneous parametric down conversion (SPDC) \cite{klyshko1970parametric,kwiat1995new}, generates photon pairs  that can be used directly, or can be converted into heralded single photons \cite{grice2001eliminating,mosley2008heralded,harder2016single,harder2013optimized}. $\chi^{(2)}$ interactions can also be used in frequency conversion (FC) \cite{kumar1990quantum,huang1992observation} to alter the spectral properties of input single photons. 

Both processes  involve shining a classical pump field onto a nonlinear material. The spectral amplitude function of the pump field (the pump function) and the function describing momentum conservation within the medium (the phase-matching function) determine a two-mode square-normalized complex amplitude function $\Psi$, known as the \emph{joint spectral amplitude} (JSA).  For SPDC, $\Psi$ is the bi-photon wavefunction of the photon pair, while for FC, $\Psi$ is the transfer function between the input and output photons \cite{ansari2018tailoring}. 

The desired form of $\Psi$  will depend on the specific situation. In Sec. \ref{sec:JSA}, we show that  if highly-pure heralded single photons are desired for high-visibility interference \cite{mosley2008heralded,harder2016single,harder2013optimized}, or if high mode-selectivity is desired for efficient frequency conversion---e.g. using a quantum pulse gate \cite{eckstein2011quantum,brecht2011quantum,brecht2015photon}---then  $\Psi$ should be \emph{separable}.  

It is well-known that, under certain conditions, Gaussian pump and phase-matching functions can make $\Psi$ separable \cite{URen2005}. This begs the question: are these the \emph{only} functions that  make $\Psi$ separable?  In Sec. \ref{nogo}, we prove that they are.

In light of this result, a second question  naturally arrises: if one of the functions is non-Gaussian, what should the other function be to maximize separability? To answer this, we develop a framework to optimize the shape of the pump  (or phase-matching) function for a given phase-matching (or pump) function, in Sec. \ref{optimalgauss}. Numerical simulations, supported by an analytical perturbative calculation, lead us to conjecture that given any phase-matching function or pump function, the optimal shape for the other function is always Gaussian.

Our proof and conjecture are derived assuming no  group velocity dispersion (GVD) between the interacting fields.  GVD is negligible for sufficiently narrow pump functions. In this limit, the Hamiltonian governing $\chi^{(2)}$ processes is equivalent to single-pump $\chi^{(3)}$ processes  \cite{quesada2014effects} (but not to dual-pump SFWM \cite{fang2013state,reddy2014efficient,Silverstone2014,Helt2017}). We comment on this in Appendix \ref{sec:dp}. The results derived here  can therefore be applied to single-pump SFWM, and to $\chi^{(3)}$ mediated frequency conversion. 

We also extend our analysis to study effects of GVD on separability. We show that, when the pump is broad,  GVD compromises separability in regions previously thought to be separable.
 
Our results also connect with the field of continuous variable quantum information, where Gaussian channels have been studied for fixed \emph{entropy} of the incoming modes \cite{guha2008entropy}. The study we perform here is the continuous variable generalization to the optimization of the entanglement of a pair of qubits under a fixed interaction \cite{dur2001entanglement}. Because we obtain our results by mapping the construction of $\Psi$ to the entanglement properties of two quantum harmonic oscillator states going through a beam splitter, we also conjecture that only a squeezed state minimizes entanglement when sent through a  beam splitter with another pre-specified ket.

 \section{The joint spectral amplitude}\label{sec:JSA}

In both SPDC and FC, the joint spectral amplitude $\Psi$ is proportional to a product of two other functions. One, known as the \emph{pump amplitude function}, captures energy conservation and has a form determined by the spectral amplitude of the pump field. The other, known as the \emph{phase-matching function},  captures linear momentum conservation and has a form determined by the nonlinear properties of the medium. 
 
For concreteness, we analyze SPDC, but later show how to map our results to  FC. The normalized JSA can be written as:
\eq{\label{biphb}
	\Psi(\wo,\wt) &= \sqrt{|r-s|} \phi(r \wo +s \wt) \gamma(\wo+\wt),
}
where $\phi$ is the phase-matching function and $\gamma$ is the pump amplitude function. A full derivation of the JSA is in Appendix \ref{sec:SPDC}. Note that $\Psi$ is unchanged if we simultaneously exchange $x \leftrightarrow y$ and $r \leftrightarrow s$, thus we will assume without loss of generality that $|r| \geq |s|$.
  The dimensionless parameters $\wo$ and $\wt$ are relative frequencies scaled by a characteristic time scale  $\tau$ (the pulse ``duration''): 
\eq{
	\wo &= \tau (\omega_{1}-\bar \omega_{1}), \quad \wt = \tau (\omega_{2}-\bar \omega_{2}),
}
where $\omega_{1/2}$ parametrize the frequency distribution of the generated photons in orthogonal modes `1' and `2', and $\bar\omega_{1/2}$ are their mean frequencies. The parameters $r$ and $s$ are also dimensionless, and depend on the length $L$ of the nonlinear material, the group velocities $v_{0/1/2}$ (where ``0'' labels the pump mode) of the fields inside the nonlinear material, and $\tau$: 
\eq{
	r &= \frac{L}{\tau}\left(\frac{1}{2 v_{1}} -\frac{1}{2 v_0} \right), \quad
	s = \frac{L}{\tau}\left(\frac{1}{2 v_{2}} -\frac{1}{2 v_0} \right).
}

The JSA in Eq. (\ref{biphb}) is  valid under the following approximations: the pump is prepared in a strong coherent state, the experiment is in the low-gain regime (higher-order photon pairs and time-ordering corrections are negligible \cite{Branczyk2011b,quesada2014effects,quesada2015time}), and  the generated photons are not too spread out around the central frequencies, i.e. group velocity dispersion (GVD) is negligible.  Eq. (\ref{biphb}) becomes zero if $r = s$, but this regime is beyond the negligible-GVD approximation.

The normalized FC transfer function can be formally obtained from Eq. (\ref{biphb}) by taking $\gamma \to  \gamma^*$ and $y \to -y$
A full derivation of the normalized transfer function is in Appendix \ref{sec:fc}.

 \subsection{Spectral purity, entanglement, and JSA separability}\label{separability}
 
SPDC photon pairs are a popular resource for heralded single photons. Detection of a photon in one mode signals the preparation of a photon in the other mode. For many applications, the heralded photons are sent to interfere---with themselves or with other photons---inside optical networks. To ensure high visibility interference, the heralded photon should be prepared in a pure quantum state \cite{Branczyk2017a}.

Detection of a single photon (using a spectrally flat photon-number-resolving detector)  in say, mode `2',  yields a heralded state in mode `1' equivalent to the marginal density matrix of the signal subsystem:
\eq{
	\rho(\wo,\wo') = \int d\wt \ \Psi(\wo,\wt) \Psi^*(\wt,\wo').
}
The spectral purity of this subsystem is  
 \eq{\label{eq:pur}
	P\left\{ \Psi \right\} = \int d\wo d\wo ' |\rho(\wo,\wo')|^2,
}
and is bounded as $0 < P\left\{ \Psi \right\} \leq 1$.  If  the bi-photon wavefunction is \emph{separable}: 
 \eq{\label{purity}
 \Psi(\wo,\wt)=\ f(\wo) g(\wt),
}
then $P\left\{ \Psi \right\}=1$. The inverse is also true. Any entanglement between the two photons reduces the purity of the heralded photon. For high-visibility interference, the bi-photon wavefunction is therefore desired to be separable. 
 
If $\Psi$ is a bi-photon wave function, the purity is a bona-fide entanglement measure, and is equal to one minus the  linear entropy \cite{buchleitner2008entanglement}. But  $P\left\{ \Psi \right\} $ characterizes the separability of \emph{any} two-dimensional function.
We therefore use the term \emph{purity} more broadly as a measure of separability for $\Psi$. The purity as defined here is a good indicator of a natural figure of merit, known as the add/drop ``selectivity'', for designing quantum pulse gates  \cite{reddy2013temporal}. 

\begin{figure}[t]
	\centering
	(a) Highly non-separable $\Psi$\\
	\includegraphics[width=0.45\textwidth]{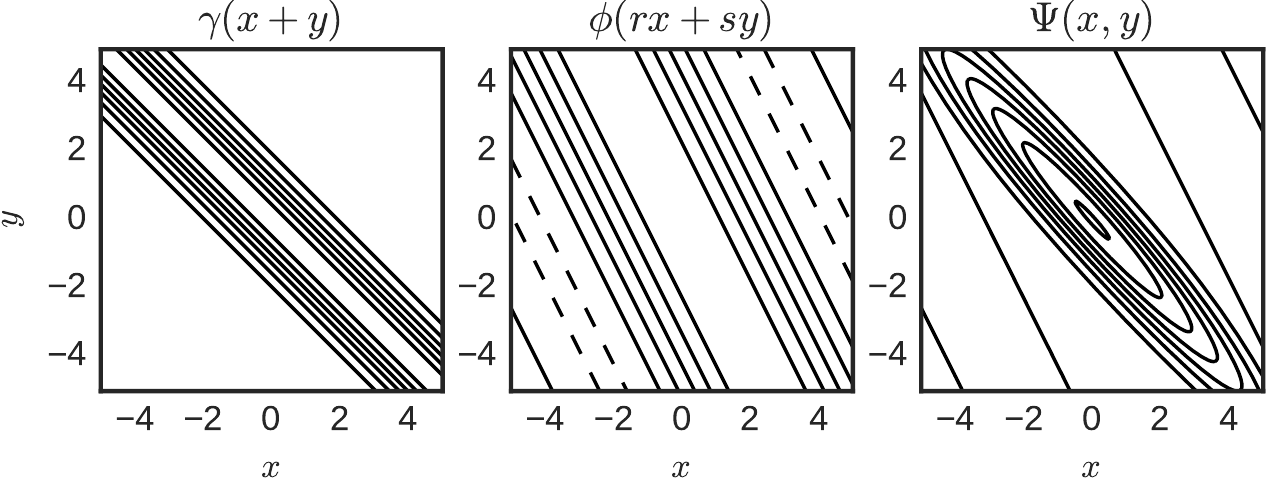}\\
	(b) Highly separable $\Psi$\\
	\includegraphics[width=0.45\textwidth]{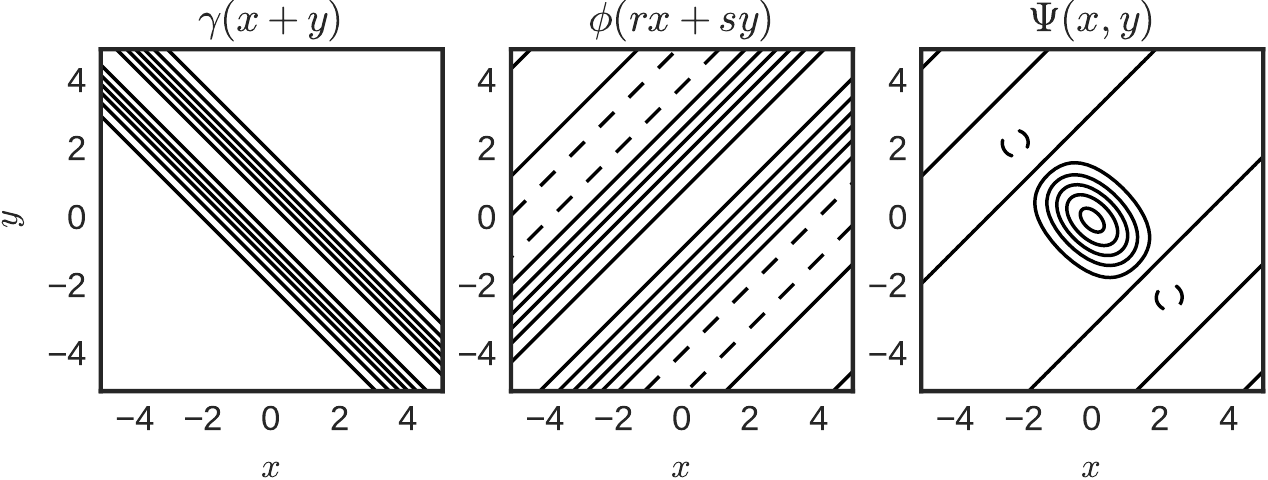}
	\caption{\label{jsaent} Generation of a biphoton wavefunction with small (top) and large (bottom) amounts of correlations.  We take the functions $\gamma(x) =  \exp(-x^2/2)/\sqrt[4]{ \pi }$ and $\phi(x) = \text{sinc}(x)/\sqrt{\pi}$. In the top figure $r =-s=1$ and  $P\{\Psi\} = 0.77$. The contour lines of the PMF are at $45^{\circ}$ with respect to the horizontal line, and   perpendicular to the contour lines of the pump function. In the bottom figure  $r = 1$, $s=1/2$ and $P\{\Psi\}= 0.24$.  The contours of the PMF  make a negative angle with respect to the horizontal line, and  the variables $x$ and $y$ are highly correlated. 
	}
\end{figure}

It is instructive to look at the contour levels of  $\gamma(\wo+\wt)$ and $\phi(r \wo+s \wt)$ (Fig. \ref{jsaent}). The contours of   $\gamma(\wo+\wt)$ are at $-45^\circ$ with respect to the vertical (angles measured in the \emph{clockwise} direction are taken to be positve), while the contours of $\phi(r \wo+s \wt)$ make an angle $\varphi$ where
\eq{\label{varphi}
\tan(\varphi) = -\frac{s}{r}
} with respect to the $\wt=0$ axis. When the pump and phasemaching functions are close to parallel, i.e. $\varphi$ is negative, $\Psi$ is highly correlated (highly non-separable) (Fig. \ref{jsaent} (a)). Since we want to work in a regime that minimizes correlations,  we can restrict our analysis to non-negative    $\varphi$ ($r s \leq 0$). Setting the angle $\varphi$, e.g.  close to $45^\circ$, can generate a less correlated (more separable) $\Psi$ (Fig. \ref{jsaent} (b)). 

In higher-gain regimes, other effects can reduce the purity of the heralded state. For example, higher-photon-pair terms can result in reduced purity in  the photon number \cite{Branczyk2010} degree of freedom. In this paper, we work in the low gain regime, and can neglect these effects.

\section{Only  Gaussians factorize the JSA: a no go result}\label{nogo}

Gaussian  $\phi$ and $\gamma$ are known to generate separable joint spectral amplitudes \cite{URen2005}. But is exact separability possible with other types of $\phi$ and $\gamma$? In this section, we show that the answer is ``no''. 

Our proof builds on results in the field of continuous variable quantum information (CVQI), where the systems of interest are quantum harmonic oscillators (QHOs). To leverage these results, we map our problem to that of two QHOs interacting on a beam splitter.

\subsection{Mapping  spectral functions to QHO states}\label{sec:map}

We first map the spectral functions in our problem to position wavefunctions of two   harmonic oscillators:
\sseq{\label{eq:phi}
	\phi(x) & \to  \ket{  \phi} = \int dx \   \phi(x) \ \ket{x}; \quad  \phi(x)=\bra{x} \phi\rangle,   \\
	\label{eq:gamma}
	\gamma(y) &\to \ket{  \gamma} = \int dy \   \gamma(y) \ \ket{y}; \quad \gamma(y)=\bra{y} \gamma\rangle,
	}
where the functions on the left of the arrow represent the pump and phase-matching functions, while the kets on the right of the arrow represent states of the QHOs, and $x$ and $y$ are dummy position integration variables. We stress that $\phi(x)$ and $ \gamma(y)$ are \emph{not} wavefunctions of the photons but wavefunction of the abstract QHO systems. The kets $\ket{x}$ and $\ket{y}$ are eigenkets of the position quadratures:
\eq{\label{position}
\hat x = (\hat a+\hat a^\dagger)/\sqrt{2}, \quad \hat y = (\hat b+\hat b^\dagger)/\sqrt{2}
}
of the QHO with destruction operators $\hat a$ and $\hat b$. A review of basic operations on bipartite QHO systems can be found in Appendix \ref{review}.

Two common operations in CVQI (also used in the field of quantum optics) are the squeezing and beam splitter operations, defined  by the operators:
\eq{
\hat S(\mu) &= \exp\left(\frac{\ln(\mu)}{2}(\hat a^2 -\hat a^{\dagger2}) \right),\\
\uu_{\text{BS}} (\theta) &= \exp\left(\theta \left( \hat a^\dagger \hat b -\hat a \hat b^\dagger  \right) \right)\,,
}
where $0<\mu =\exp(r)$, and $r  \in \mathbb{R}$ is the usual squeezing parameter and a completely analogous definition of the squeezing operator exists for mode $\hat b$. We parametrize squeezing by the exponential of the ``usual'' squeezing parameter which corresponds to how much the $x$ axis is squeezed $\mu<1$ or  anti-squeezed $\mu>1$. The parameter $\theta$ is related to  the beam splitter reflectivity.

\begin{figure}[t]
	\centering
	\includegraphics[width=0.47\textwidth]{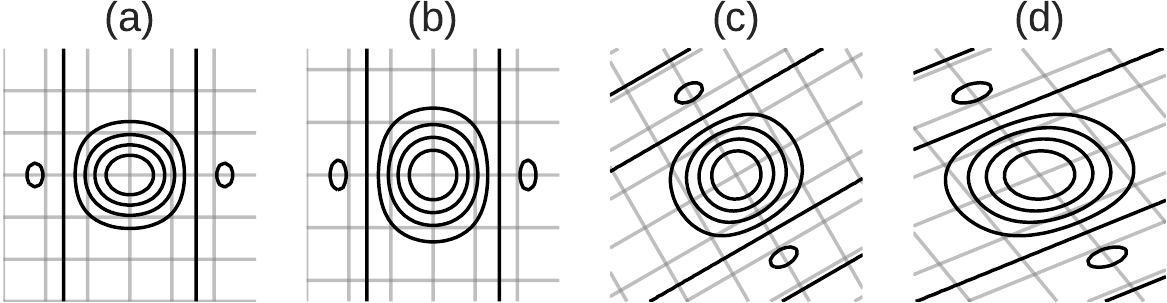}
	\caption{\label{cart} A graphical representation of the linear transformations: (a) $\ket{\Psi'''} = \ket{\phi} \otimes \ket{\gamma} 	\to$ (b)	$ \ket{\Psi''} = \mathbb{ \hat I} \otimes \hat S(\nu) \ket{\Psi'''} \to$ (c) $
	 \ket{\Psi'} =  \mathcal{\hat U}_{\text{BS}}(\theta) \ket{\Psi''}\to$ (d) $
	 \ket{\Psi} = \hat S(\kappa) \otimes \hat S(\sigma) \ket{\Psi'}$. 
}
\end{figure}

Now imagine  a system prepared in the state  $\ket{\gamma}$  from Eq. (\ref{eq:gamma}), then  squeezed, then sent to interact on a beam splitter with  another system prepared in the state $\ket{\phi}$ from Eq. (\ref{eq:phi}). Upon exiting, each system gets squeezed again. The output state is
\eq{\label{paramet}
	\ket{\Psi}=&   \left( \hat S(\kappa) \otimes \hat S(\sigma) \right) \mathcal{\hat U}_{\text{BS}}(\theta) \left(  \mathbb{\hat I} \otimes \hat S(\nu) \right) \ket{\phi} \otimes \ket{\gamma}\,.
}
The successive operations of squeezing, rotation and skewing (squeezing in a rotated basis),
are shown sequentially in panels (a)$-$(d) of Fig. \ref{cart}.

We again stress that we are not modelling actual photons being squeezed and sent through a physical beam splitter, but rather applying abstract squeezing and beamsplitter operations to abstract states of the QHO.

Now consider the squeezing parameters:
\eq{\label{solall}
	\kappa = \sqrt{r (r -s)};~~\sigma = \sqrt{s (s -r)};~~\nu = \frac{1}{\sqrt{-r s}}\,,
}
and  beam splitter parameter:
\eq{
	\tan(\theta) &= \frac{\sigma}{\kappa} = \sqrt{-\frac{s}{r}} = \sqrt{\tan{\varphi}}\,,
}
where we assumed $rs < 0 $ and $\phi$ is the angle the contours of the PMF make with the vertical axis (See Eq. (\ref{varphi})). 
For this choice of parameters, the position wavefunction of the ket in Eq. (\ref{paramet}) $(\bra{x} \otimes \bra{y} ) \ket{\Psi}$ is precisely the JSA
\eq{
\Psi(x,y)=\sqrt{|r-s|}\phi(r x+ sy) \phi(x+y).
}
This exact equivalence is shown in Appendix \ref{proof22}.

\subsection{Separability proof}

Using the mapping in the previous section, we have  shown that the entanglement of the JSA is equivalent to the entanglement acquired by  the state $\ket{\phi} \otimes \hat S(\nu) \ket{\gamma}$ when sent through a beam splitter $\uu_{\text{BS}}(\theta)$. A beam splitter is a 2-port linear-optical operation,  so we can use the following elegant result for $N$-port linear-optical networks by Jiang, Lang and Caves \cite{jiang13}:
\begin{lemma}\label{th2}
	Given a nonclassical pure-product-state input to an $N$-port linear-optical network, the output is almost always mode entangled; the only exception is a product of squeezed states, all with the same squeezing strength, input to a network that does not mix the squeezed and anti-squeezed quadratures.
\end{lemma}
An immediate consequence of this result is that to have no entanglement at the output of the beam splitter, it is necessary that the input state to the beam splitter is a product of two squeezed states, squeezed by the same amount. This implies that  the states of the two harmonic oscillators must be Gaussian in the position basis. According to the mapping in Sec. \ref{sec:map},  this in turn implies that for the JSA to be separable, the PMF and pump function \emph{must} be Gaussian. 

To determine which Gaussian functions guarantee separability, consider phase-matching and pump functions of the form
\seq{
	\label{2gaussians}
\eq{
	\phi(r \wo+s \wt) &= \frac{1}{\sqrt[4]{ \pi}} e^{-\frac{1}{2}(r \wo+s \wt)^2},\\
	\gamma(\wo+\wt) &= \frac{1}{\sqrt[4]{ \pi}} e^{-\frac{1}{2} (\wo+\wt)^2}.
}
}
According to Eq. (\ref{biphb}), the biphoton wavefunction is 
\eq{\label{separable}
	\Psi(\wo,\wt)&=\frac{\sqrt{|r-s|}}{\sqrt{  \pi}} \exp\left(-\frac{1}{2} \left[\wo \  \wt \right] \mathbf{C}
	\left[
	\begin{array}{c}
		\wo \\
		\wt\\
	\end{array}
	\right]
	\right),
}
where 
\eq{\label{c}
	\mathbf{C}&=\left[
	\begin{array}{cc}
		1+r^2 & 1+r s \\
		1+r s & 1+s^2 \\
	\end{array}
	\right],
}
is the correlation matrix. The function $\Psi(\wo,\wt)$ is separable iff the off-diagonal elements of the correlation matrix are zero, i.e. if $rs=-1$ (consistent with \cite{URen2005}). This proves that:
\begin{theorem}\label{th1}
	Given a pump function $\gamma$ and a phase-matching function $\phi$,  the joint spectral amplitude $\Psi(\wo,\wt) = \sqrt{|r-s|} \phi(r \wo +s \wt) \gamma(\wo+\wt)$ is  almost never separable in $\wo$ and $\wt$; the only exception is when $\gamma$ and $\phi$ are Gaussian functions as in Eq. (\ref{2gaussians}), and $r s = -1$.
\end{theorem}
\begin{figure}[t]
	\centering
	\includegraphics[width=0.47\textwidth]{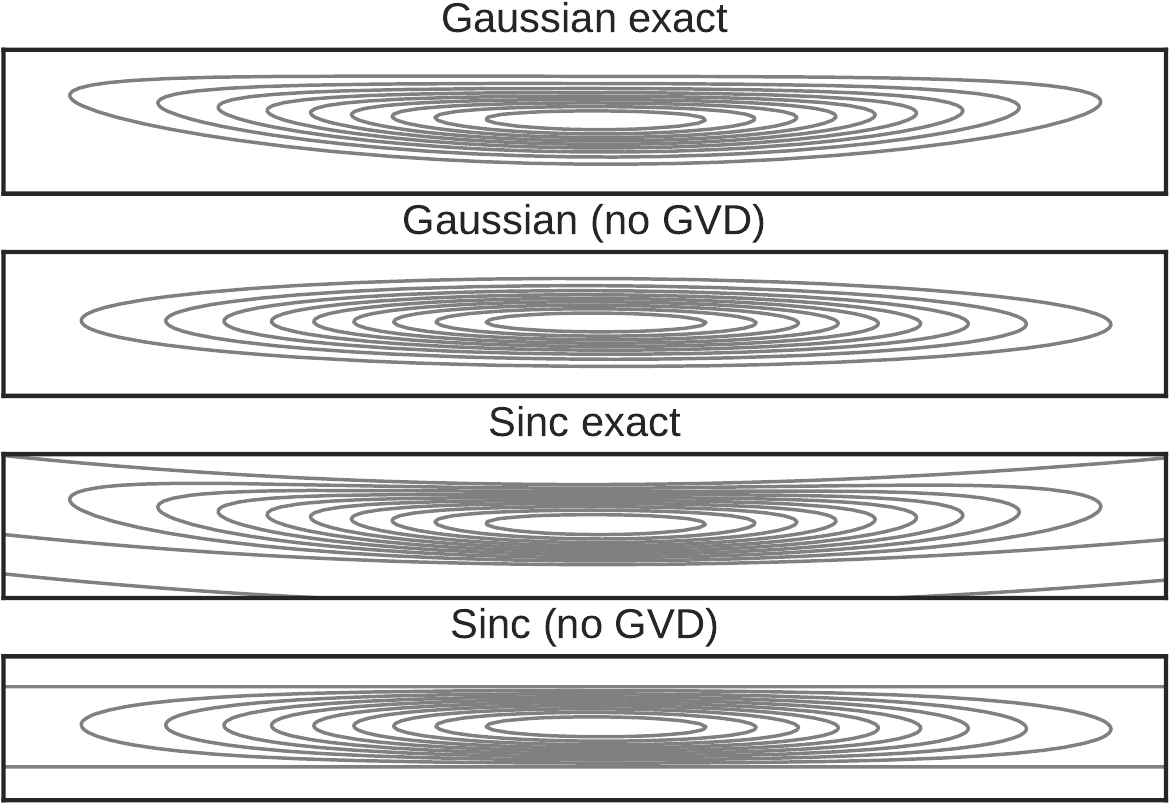}
	\caption{\label{JSAxx} JSA for Gaussian and sinc  PMFs, with and without  group velocity dispersion (GVD), for $r= 23.4$ and $s=0$. Notice that GVD introduces curvature in the JSA. We used  Sellmeier equations for Potassium Titanyl Phosphate (KTP) \cite{Sellmeier2018}, with crystal length $L=0.02$ m, pumped by a $1211$ nm Gaussian pump with $\tau=2.9\times10^{-14}$ s.  The horizontal and vertical frequency ranges are   $[7.5,8.1]\times10^{14}$ s$^{-1}$ and $[7.7,7.8]\times10^{14}$ s$^{-1}$ respectively. Note that in this graph the $x$ axis is vertical and the $y$ is horizontal}
\end{figure}

When $r =1$ and $s=-1$, the pump and phase-matching functions  are perpendicular and equally wide. This regime is known as \emph{symmetric group velocity matching} and generates separable photons with equal bandwidths. But the more general condition $rs=-1$ means that exact separability is achievable for any values of the PMF angle $\varphi$ which, once $r s = -1$ is imposed, is simply equal to $\varphi = \tan^{-1}(s^2)$.

The results for SPDC can be mapped to frequency conversion. In Appendix \ref{sec:fcmap}, we show that the entanglement of the bi-photon wave function and the frequency conversion transfer function is the same (modulo the substitution of $\gamma$ by $\gamma^*$).

\subsection{Asymptotic separability}\label{sec:examp}
It is possible for the JSA to  approach separability asymptotically for any shaped pump and phase-matching functions \cite{URen2005}.  Before we consider arbitrary shapes, it is instructive to  identify this regime for Gaussian functions. 

For Gaussian pump and phase-matching functions, the analytical expression for the JSA purity is:
\eq{
	P\left\{ \Psi \right\} = \frac{\left| r-s\right|
	}{\sqrt{\left(1+r^2\right)
		\left(1+s^2\right)}}.
}
When $s=0$ this becomes
\eq{
	P\left\{ \Psi \right\}=\frac{1}{\sqrt{1+\left(\frac{1}{r}\right)^2}}\,.
}
This expression is strictly less than 1, but approaches arbitrarily close to  1 when $r\gg 1$. In this limit  
\eq{\label{scaling}
	P\left\{ \Psi \right\}\approx 1-1/(2 r^2).
} 

For arbitrary functions other than Gaussian, one has to be careful when taking the limit $s \to 0$, as it can lead to divergences in Eq. (\ref{solall}). We show how to take this limit properly in the case of $\theta\ll1$  in Sec. \ref{assym}.

\begin{figure}[t]
	\centering
	\includegraphics[width=0.47\textwidth]{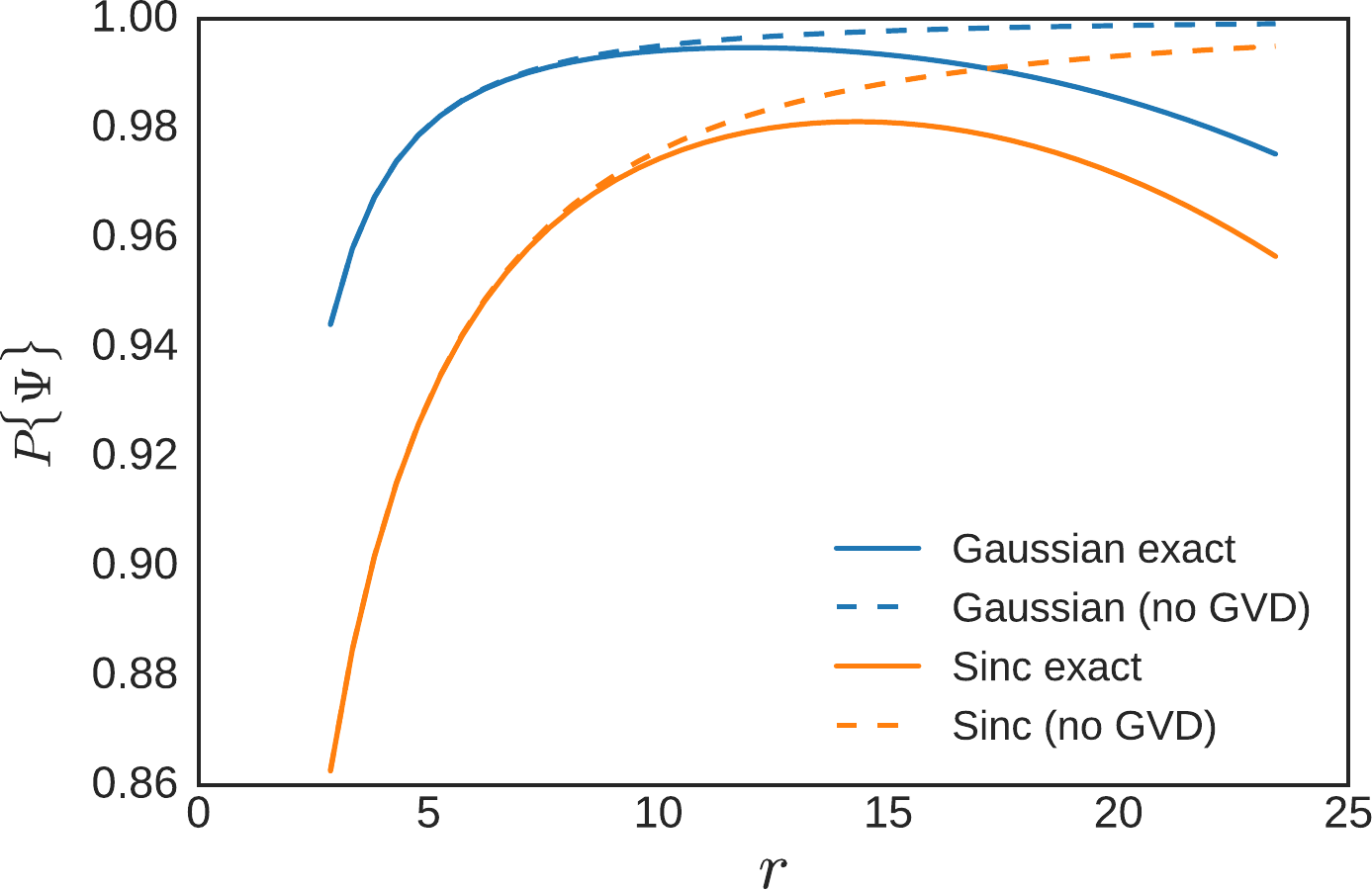}
	\caption{\label{purities} Purities for Gaussian and sinc  PMFs, with and without group velocity dispersion (GVD), as a function of $r$ (with $s=0$). All curves were computed numerically, but the dashed blue line also  corresponds to $1-1/(2 r^2)$. We used  Sellmeier equations for Potassium Titanyl Phosphate (KTP) \cite{Sellmeier2018}, for a crystal of length $L=0.02$m, and a pumped by a $1211$ nm Gaussian pump.}
\end{figure}

One also has to be mindful of the  pump function width in this regime. For wide pumps, GVD cannot be neglected, and will introduce curvature in the PMF, as shown in Fig. \ref{JSAxx}. When GVD is taken into account, the purity initially increases with $r$, but then reaches a maximum and drops, as shown in Fig \ref{purities}. 

\section{One Gaussian is better than none}\label{optimalgauss}

In the previous section, we showed that exact separability can only be achieved with Gaussian pump and  phase-matching functions.  But what if  the phase-matching function is non-Gaussian and cannot be changed? Can we find the optimal pump amplitude function $\gamma_{\mathrm{opt}}$ that maximizes the heralded-photon purity? In other words, can we find
    \eq{
    \gamma_{\mathrm{opt}}=	\max_{\gamma} P\left\{ \Psi \right\} ?
    }
    
To answer this question, we follow the same procedure as before, and map the construction of the JSA to the interaction of  two QHO states going through a beam-splitter. The optimization problem can be written as
\eq{\label{optket}
	\ket{\gamma_{\text{opt}}} = \arg \max_{\gamma} P\left\{\uu_{\text{BS}}(\theta) \left(\ket{\phi} \otimes S(\nu)\ket{\gamma}\right) \right\}
}
for fixed $\ket{\phi}$, $\theta$ and $\nu$, where $\arg \max$ is the value that maximizes the target function. The expression does not depend on $ \hat S(\kappa) \otimes \hat S(\sigma) $ because local operations (e.g. squeezing) applied \emph{after} an entangling operation (e.g. a beam splitter) do not change the amount of entanglement.  The problem can be simplified to
\eq{\label{optprob}
	\ket{\gamma_{\text{opt}}} &= \hat S\left(\nu^{-1}\right)\arg \max_{\gamma'} P\left\{ \uu_{\text{BS}}\left(\theta\right) \left(\ket{\phi} \otimes \ket{\gamma'} \right) \right\},
}
where
\eq{
	\ket{\gamma'} &= \hat S(\nu)\ket{\gamma}\,.
}

What if, on the other hand,   the pump amplitude function is non-Gaussian and cannot be changed? Can we find the optimal phase-matching function $\phi_{\mathrm{opt}}$ that maximizes the heralded-photon purity? The optimization problem becomes
\eq{\label{optprob2}
	\ket{\phi_{\text{opt}}} = \arg \max_{\phi} P\left\{ \uu_{\text{BS}}\left(\theta\right) \left(\ket{\phi} \otimes \ket{\gamma'} \right) \right\}\,.
}
This optimization problem is completely analogous to the one in (\ref{optprob}). We therefore restrict the rest of our analysis to the former.

\subsection{Picking a basis for the problem}

In Section \ref{nogo}, we framed the problem in the position basis of the QHOs, but this basis may not be ideal for numerical optimization. Here, we use the truncated number basis which provides a natural basis to represent states with a non Gaussian wavefunction.

To prepare the problem for numerical simulation, we  rewrite it as
\eq{\label{cost1}
	\max_{\gamma'} F(\gamma') = \max_{\gamma'} \frac{P\big\{\uu_{\text{BS}}(\theta) \left(\ket{\phi} \otimes \ket{\gamma'}\right) \big\} -\lambda |\braket{\gamma'|\hat b|\gamma'}|^2}{|\braket{\gamma'|\gamma'}|^2} \,,
}
where  recall that $\hat{b}$ was introduced in Eq. (\ref{position}). The term $-\lambda |\braket{\gamma'|\hat{b}|\gamma'}|^2$ is added in the function to favour kets that have zero mean displacement, since without loss of generality one can assume that  $\ket{\phi}$ and $\ket{\gamma'}$ have zero displacement
\footnote{
This is due to the fact that 
$
\uu_{\text{BS}}(\theta) \hat D(\alpha )\otimes \hat D(\beta) = \hat D(\hat \alpha')\otimes \hat D(\beta')\uu_{\text{BS}} (\theta) 
$
where the displacement operators are defined by
$
\hat D(\alpha) = \exp(\alpha \hat a^\dagger - \alpha^* \hat a), \quad
\hat D(\beta) = \exp(\beta \hat b^\dagger - \beta^* \hat b).
$
}, i.e. $
	 \braket{\hat a}_{\phi} =0, \quad 	\braket{\hat b}_{\gamma'} =0$. This restriction places most of the support of the ket in the lowest $N$ number states in the decomposition, which makes the computation more tractable. 
	 
	 The sought after ket $\ket{\gamma'}$ can be parametrized in terms of real numbers $u_n$ and $v_n$:
\eq{\label{expansion}
	\ket{\gamma'} = \sum_{n=0}^{N-1} (u_n+i v_n) \ket{n}\,,
}
where $\ket{n} = (\hat b^{\dagger n})/(\sqrt{n!}) \ket{0}$. This can also be done for the fixed incoming PMF ket $\ket{\phi}$.

The denominator in Eq. \ref{cost1} makes $F(\gamma')$  insensitive to re-scaling of $\ket{\gamma'}$ thus we need not worry about its normalization. 

To perform the optimization in practice, the number-state-decompositions  for $\ket{\gamma'} $ and $\ket{\phi}$ are  inserted into Eq. (\ref{cost1}), and the trace is taken in the number basis. The expression is then optimized numerically over the real parameters $u_n$ and $ v_n$.

\subsection{Numerical optimization}\label{numopt}

	By using the automatic differentiation capabilities of the tensorflow \cite{abadi2016tensorflow} backend of strawberryfields \cite{killoran2018strawberry} we implemented a gradient-based optimization in the truncated ($N=30$) number basis using . We considered the following phase-matching function
\eq{\label{sinc}
	\phi(x)\rightarrow\phi(x)=\braket{x|\phi} = \frac{\text{sinc}\left(\frac{x}{\alpha}\right)}{\sqrt{\alpha \pi}} \,,
	}
with $\alpha \approx 0.71$. This value was chosen for convenience so as to maximize the overlap 
\eq{
  \int dx \frac{1}{\sqrt[4]{ \pi}} e^{-\frac{x^2}{2}} \phi(x) = \sqrt[4]{\pi } \sqrt{\alpha } \ \text{erf}\left(\frac{1}{\sqrt{2} \alpha }\right)\,,
}
thus maximizing the amplitude of  $\ket{\phi}$ in the 0 energy state of the QHO, which is helpful in capturing most of the state $\phi$ in Eq. (\ref{sinc}) in the lowest $N$ number states that we use to represent the state.

\begin{figure}[t]
	\centering
	\includegraphics[width=0.47\textwidth]{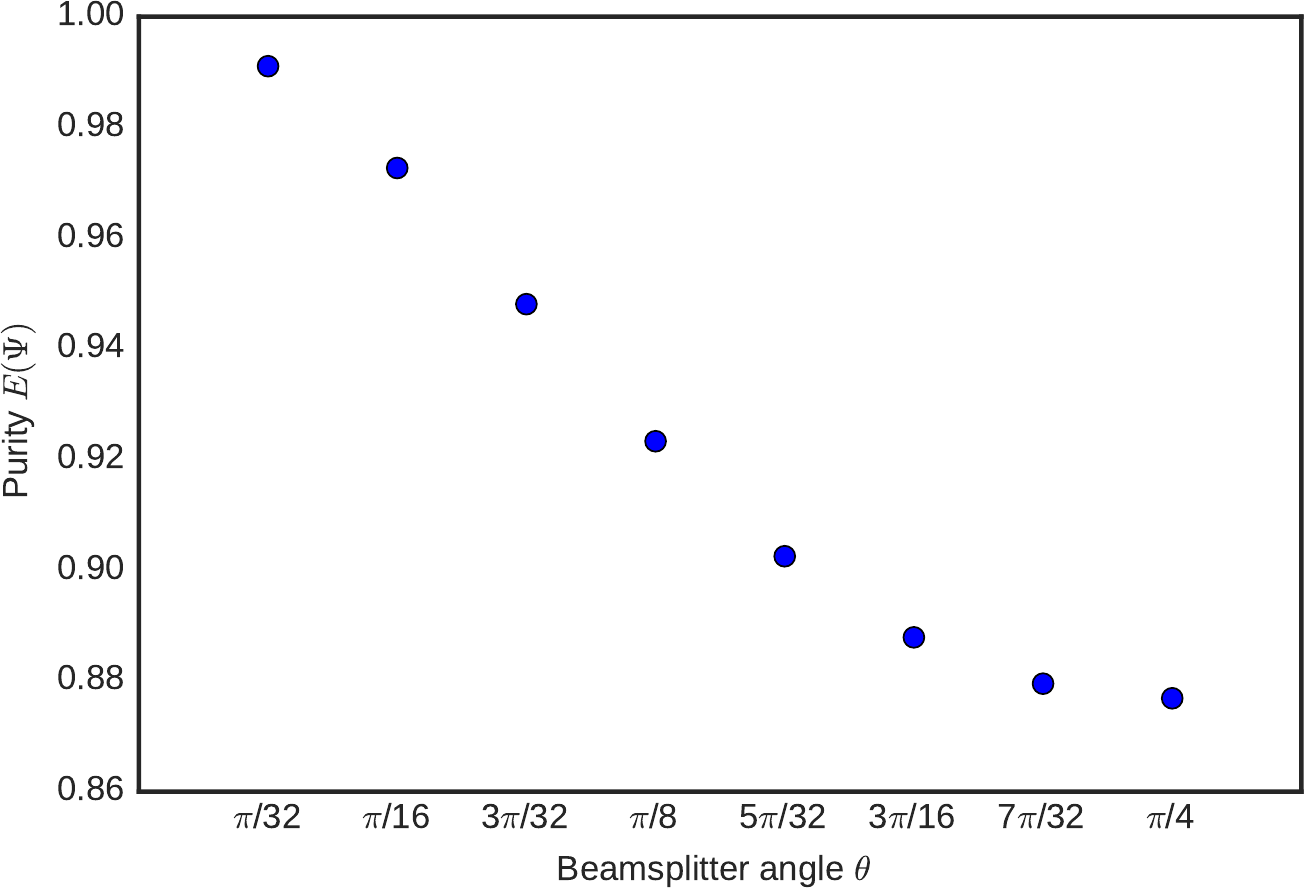}
	\caption{\label{opts} Numerical optimization results for the input sinc function in Eq. (\ref{sinc}). In all cases the highest purity plotted in this figure was achieved by a state with a fidelity of at least 99.9 \% with a squeezed vacuum state.}
\end{figure}

Using the cost function in Eq. (\ref{cost1}) and a global optimization procedure which consists of random generation of kets of the form in Eq. (\ref{expansion}) (of which we considered 80 candidates per value of $\theta$), followed by gradient descent, we found that for $\theta \in \{ k \pi/32\}_{k=1}^8$ the optimal pump ket $\ket{\gamma'}$ always had a fidelity of at least $99.9 \%$ with a squeezed state. The optimal purities for the different values of $\theta$ are shown in Fig. (\ref{opts}).  This tells us that for any $\theta$ in Eq. (\ref{solall}), and therefore for any angle $\varphi$ between the pump and sinc-shaped phase-matching function, the optimal pump function is Gaussian.

\subsection{Analytical optimization for $\varphi, \theta \ll 1$}\label{smalltheta}

The analytical and numerical results presented so far have shown that for Gaussian and sinc PMFs, the optimal pump function is always a Gaussian function. Is a Gaussian pump function optimal for \emph{arbitrary} PMFs? In this section, we analytically prove that,  for $\theta \ll 1$,  it is. We do so by showing that, in the limit $\theta \ll 1$,  the state that maximizes the purity in Eq. (\ref{optprob}) is a squeezed state. 

In the limit $\theta \ll 1$, the purity is
\eq{\label{eq:lim}
	P\left\{ \Psi \right\}&\approx
	1-(2\theta)^2\left(n_{\phi} n_{\gamma'} +\frac{n_{\gamma'}+n_{\phi}}{2} - \Re[m^*_{\gamma'} m_\phi] \right),
}
where $n_{\phi}=\braket{\hat a^\dagger \hat a}_{\phi}$ and $n_{\gamma'}=\braket{\hat b^\dagger \hat b}_{\gamma'}$, $m_{\phi}=\braket{\hat a^2 }_{\phi}$ and $m_{\gamma'}=\braket{\hat b^2 }_{\gamma'}$, and we assumed that $\braket{\hat a}_{\phi} = \braket{\hat b}_{\gamma'} =0$ without loss of generality. The expression in Eq. (\ref{eq:lim}) is derived in  Appendix A of  Ref. \cite{goldberg2017nonclassical} (where their rotation angle is half that used here, hence the factor of $2 \theta$).

We consider a fixed state $\phi$ in mode $\hat a$, and therefore fix $n_\phi,m_\phi$. Maximizing the purity  in Eq. (\ref{eq:lim}) is equivalent to minimizing the quantity inside the round brackets on the right hand side of Eq. (\ref{eq:lim}), which in turn is equivalent to maximizing the first two terms in the round brackets while maximizing $\Re[m^*_{\gamma'} m_\phi] $. The last term $\Re[m^*_{\gamma'} m_\phi] $ is maximum when $m^*_{\gamma'} m_{\phi}$ is positive, i.e., when  $\arg(m_\phi)=\arg(m_{\gamma'})$. We therefore fix the phase of $m_{\gamma'}$ to equal that of $m_{\phi}$, and maximize the quantity
\eq{
n_{\phi} n_{\gamma'} +\frac{n_{\phi}+n_{\gamma'}}{2} - |m_{\phi}||m_{\gamma'}| .
}
To minimize the quantity inside the brackets we want to make $n_{\gamma'}$ as small as possible while making $m_{\gamma'}$ as large as possible. For a given $n_{\gamma'}$ the following expression, equivalent to the uncertainty principle, is satisfied:
\eq{\label{heis}
	|m_{\gamma'}| \leq \sqrt{n_{\gamma'} (n_{\gamma'}+1)}\,,
} with  equality attained only for squeezed states \cite{lang2013optimal,lang2014optimal}. If squeezed states maximize $|m_{\gamma'}|$, they minimize the quantity inside the brackets in Eq. (\ref{eq:lim}), and thus maximize the purity.

We now want to see which squeezed state this is. Using the optimal value for $|m_{\gamma'}| = \sqrt{n_{\gamma'}(n_{\gamma'}+1)}$, the purity is
 \eq{
	P\left\{ \Psi \right\}&\approx
	1-(2\theta)^2 \times \\
	&\left(n_{\phi} n_{\gamma'} +\frac{n_{\phi}+n_{\gamma'}}{2} - \sqrt{n_{\gamma'} (n_{\gamma'}+1)}|m_{\phi}| \right). \nonumber
}
The optimal value for $n_{\gamma'}$ can be found, by finding the $n_{\gamma'}$ such that $\frac{d  P\left\{\Psi \right\} }{d n_{\gamma'}}=0$, to be
\eq{\label{nopt}
	n_{\gamma'}^{\text{opt}}&\to \frac{1}{8 \left| m_{\phi}\right| ^2-2 \left(2n_{\phi}+1\right)^2}   \nonumber \\
&	\times\Bigg(1+4 n_{\phi} \left(n_{\phi}+1\right) -4 \left|m_{\phi}\right|^2 \Bigg. \nonumber \\
&	\Bigg. -\sqrt{\left(2 n_{\phi}+1\right){}^2 \left(\left(2 n_{\phi}+1\right)^2-4\left| m_{\phi}\right| {}^2\right)} \Bigg)\,.
}
In the limit where $|m_{\phi}| \to \sqrt{n_{\phi} (n_{\phi}+1)}$ one has $n_{\gamma'} \to n_{\phi}$ and recovers the well-known result for a pair of squeezed states.
The quantity $n_{\gamma'}$ is related to the squeezing parameter $r = \ln(\mu)$ of a squeezed vacuum state as $\sinh^2 (r) = n_{\gamma'}$.

We have therefore shown that, in the limit $\theta \ll 1$,  a squeezed vacuum state with  parameters given in Eq. (\ref{nopt}) minimizes entanglement when sent through a fixed beam splitter with another pre-specified ket. 

Using our mapping, this means that, in the limit $\theta \ll 1$,  Gaussian pump or phase-matching functions maximize JSA separability when the other function is non-Gaussian.

\subsection{Asymptotic separability}\label{assym}

In Section \ref{sec:examp}, we considered a regime where the JSA approaches separability asymptotically for Gaussian pump and phase-matching functions. For arbitrary functions,  one needs to be careful when taking the limit $s \to 0$, as it can lead to divergences in Eq. (\ref{solall}). 

In this section, we take advantage of the results for the  limit $\theta \ll 1$ from  Section \ref{smalltheta}  to look at the GVM limit $s \to 0$ and confirm that if GVD can be neglected and $r \gg 1$ one can approach separability for arbitrary shaped functions. However for long enough pulses GVD is always a problem as shown in Figs.  (\ref{JSAxx}) and (\ref{purities}).

The purity in Eq. (\ref{eq:lim}) is written in terms of the moments $n_{\gamma'}, m_{\gamma'}$ of the state $ \ket{\gamma'}=\hat S(\nu) \ket{\gamma}$, which we now write in terms of the moments of the state $\ket{\gamma}$:
\eq{\label{eq:mom}
	n_{\gamma'}=&-\frac{1}{4} \nu ^2 m_{\gamma }{}^*+\frac{m_{\gamma }{}^*}{4 \nu ^2}+\frac{\nu
		^2}{4}+\frac{1}{4 \nu ^2}-\frac{\nu ^2 m_{\gamma }}{4}\nonumber \\
	&+\frac{m_{\gamma }}{4 \nu
		^2}+\frac{\nu ^2 n_{\gamma }}{2}+\frac{n_{\gamma }}{2 \nu ^2}-\frac{1}{2} ;\\
	m_{\gamma'}=&\frac{\nu ^2 m_{\gamma }{}^*}{4}+\frac{m_{\gamma }{}^*}{4 \nu ^2}-\frac{m_{\gamma
		}{}^*}{2}-\frac{\nu ^2}{4}+\frac{1}{4 \nu ^2}+\frac{\nu ^2 m_{\gamma
	}}{4}\nonumber \\
	&+\frac{m_{\gamma }}{4 \nu ^2}+\frac{m_{\gamma }}{2}-\frac{\nu ^2 n_{\gamma
}}{2}+\frac{n_{\gamma }}{2 \nu ^2}\,.
	}

From Eq. (\ref{solall}), we see that taking the limit $s \to 0$ amounts to taking the limits $\theta \to 0$ and $\nu \to \infty$. Naively, one might think that taking the limit $\nu \to \infty$ would cause the moments in Eq. (\ref{eq:mom}) to diverge. But since $\theta \nu \to 1/r \ll 1$, we obtain
\eq{\label{limitfinal}
\begin{split}
	P\left\{ \Psi \right\} \approx {}&1-\frac{1}{2 r^2}  \left(1-2 \Re (m_{\gamma})+2 n_{\gamma}\right) \\
	&\times\left(1+2 \Re(m_{\phi})+2 n_{\phi}\right)\,.
	\end{split}
	}
Each term inside round brackets is positive. This is  established by combining Eq. (\ref{heis}) with the arithmetic mean-geometric mean inequality to obtain
\eq{
\begin{split}
	|\Re(m_{\gamma/\phi})| &\leq{} |m_{\gamma/\phi}| \\
	&\leq \sqrt{n_{\gamma/\phi}(n_{\gamma/\phi}+1)} < \frac{2 n_{\gamma/\phi} +1}{2}\,.
	\end{split}
	}
This confirms that while exact JSA separability cannot be reached, it can be approached arbitrarily close. We saw this in closed form for Gaussians at the end of Sec. \ref{sec:examp}, but we see now that the scaling in Eq. (\ref{scaling}) is completely general for any PMF and pump function (in the limit $\theta\ll 1$) as can be seen for e.g. a sinc PMF in Fig. \ref{purities}.

\subsection{Conjecture}

At the beginning of this section, we asked the question: if one of the functions is non-Gaussian, what should the other function be to maximize separability? To answer this, we developed a framework to optimize the shape of the pump (or phase-matching) function for a given phase-matching (or pump) function. 

This framework did not give a conclusive answer, but provided strong evidence for us to make a conjecture. Before presenting the conjecture, we summarize our results:
\begin{itemize}
\setlength\itemsep{0em}
\item Given a Gaussian phase matching function $\phi$,  the JSA is exactly separable only when the pump function $\gamma$ is a Gaussian function (and vice-versa).  Equivalently, given a squeezed state $\ket{\phi}$ incident on a beam splitter with the state $\ket{\gamma'}$, the output is unentangled only when  $\ket{\gamma'}$ is a squeezed state, squeezed by the same amount as $\ket{\phi}$. 
\item  When $s/r\ll 1$, given any non-Gaussian phase matching function $\phi$, we showed analytically that the JSA is most separable when the pump function $\gamma$ is  a Gaussian function (and vice-versa). Equivalently, When $\theta \ll 1$, given any arbitrary state $\phi$ incident on a beam splitter with the state $\ket{\gamma'}$, we showed analytically that the output is least entangled when  $\ket{\gamma'}$ is a squeezed state.
\item For values away from $s/r\ll 1$, given a highly non-Gaussian, i.e. sinc, phase matching function $\phi$, we showed numerically that the JSA is most separable when   the pump function $\gamma$ is  a Gaussian function (and vice-versa). Equivalently, for values away from $\theta \ll 1$, given a state $\ket{\phi}$  with a highly non-Gaussian, i.e. sinc, wavefunction, incident on a beam splitter with the state $\ket{\gamma'}$, we showed numerically that the output is least entangled when  $\ket{\gamma'}$ is a squeezed state.
\end{itemize}
These three results lead us to the following conjecture:
\begin{conjecture}
To maximize JSA separability when one of the (pump or phase matching) functions is non-Gaussian, the other function \emph{must} be Gaussian.
Or in terms of our optimization problem:
The solution to the problem 
\eq{
 \arg \max_{\gamma'} P\left\{ \uu_{\text{BS}}\left(\theta\right) \left(\ket{\phi} \otimes \ket{\gamma'} \right) \right\}
}
for fixed $\theta$ and $\ket{\phi}$ is always a squeezed state.
\end{conjecture}

\section{Discussion}

Many nonlinear optical technologies such as heralded single-photon generation and frequency conversion require the joint spectral amplitude  (JSA) to be separable. It is known that, under certain conditions, the JSA factorizes when the incident pump field and nonlinear medium are described by Gaussian functions \cite{URen2005}. We proved here that this is the \emph{only} way to factorize the JSA.

Gaussian pump  and phase-matching functions (that describe the nonlinear material) can be realized in practice. While standard pulsed lasers have sech-shaped spectral amplitudes, they can be shaped using optical pulse shaping \cite{Weiner2011}. While standard nonlinear materials yield sinc-shaped phase-matching functions, they can be shaped using custom-poling methods in $\chi^{(2)}$ materials \cite{Branczyk2011,Dixon2013, Dosseva2014,Tambasco2016,Graffitti2017,Graffitti2018}, and pulse-delay methods in $\chi^{(3)}$ materials \cite{fang2013state}.

For situations where resources constrain the  pump field or the nonlinear medium, we developed a framework to optimize the shape of the pump  (or phase-matching) function for a given phase-matching (or pump) function. Numerical simulations, supported by an analytical perturbative calculation, lead us to conjecture that given any phase-matching function or pump function, the optimal shape for the other function is always Gaussian.

But our optimization framework has potential for broader applicability. It can be adapted to include features such as chirped pulses to compensate for temporal broadening, or incorporate variations in nonlinear properties along the material. It can also be adapted to optimize the JSA for situations other than separability, such as generating photons with interesting temporal shapes, as was done using FC in  \cite{ansari2018tailoring} (using other methods),  optimizing for frequency comb generation, as was done in \cite{arzani2018versatile} (using other methods), or preparing hyper-entangled photon pairs (e.g. with spectral and polarization entanglement) \cite{Kwiat1997}. 

We obtained our results by mapping the construction of the JSA to the entanglement properties of two quantum harmonic oscillator states going through a beam splitter. This mapping opens up avenues for connecting problems in continuous variable quantum information to  problems in  the field of nonlinear optics. Known results for mulit-port beam splitters might be applicable for modelling non-linear optical interactions with more modes, such as dual-pump SFWM. 
Our results on optimizing JSA separability are of practical interest to researchers using waveguide (or other quasi-1D geometry) nonlinear optics for high-quality generation and manipulation of quantum-light-sources for emerging optical technologies.

\section*{Acknowledgements}
We thank Francesco Graffitti, Alessandro Fedrizzi, and J\'er\'emy Kelly-Massicotte for providing valuable feedback on our manuscript. 

Research at Perimeter Institute is supported by the Government of Canada through Industry Canada and by the Province of Ontario through the Ministry of Research and Innovation.

\bibliographystyle{apsrev}
\bibliography{OptimizedPump} 

\newpage

\onecolumngrid
\appendix

\section{Modelling quantum states and processes mediated via $\chi^{(2)}$ processes}
Consider a $\chi^{(2)}$ nonlinear material in a quasi-one-dimensional geometry. The material mediates the mixing of three different modes: $i =\{0,1,2\}$. These modes have central frequencies $\bar \omega_i$. For any frequency $\omega_i$ in mode $i$ the dispersion relation is
\eq{\label{disp1}
k_i(\omega_i)= \frac{n_{i}(\omega_i)\omega_i}{c} \,,
}
where $n_{i}(\omega_i)$ is the refraction index of mode $i$ and $c$ is the speed of light in vacuum. 

The $\chi^{(2)}$  process is most efficient when it satisfies energy and momentum conservation according to:
\sseq{
	\bar \omega_{0}- \bar \omega_{1}+\bar \omega_{2}   &=0~~~\mathrm{and} ~~~k_{0}(\bar \omega_{0})-k_{1}(\bar \omega_{1})- k_{2}(\bar\omega_{2})=0\,.
} 
In periodically-poled materials with spatial periodicity $\Lambda$, the corresponding phase-matching condition is:
\sseq{
	k_{0}(\bar \omega_{0})-k_{1}(\bar \omega_{1})- k_{2}(\bar\omega_{2}) &=\pm \frac{2 \pi}{\Lambda}\,.
} 

For photons not too spread out around their central frequencies, one can ignore quadratic and higher terms in the group velocity dispersion. The dispersion relation in Eq. (\ref{disp1}) is then  approximated as:
\eq{\label{disp}
	k_i(\omega_{i}) = k_i(\bar \omega_{i})+\frac{(\omega_{i}-\bar \omega_{i})}{v_{i}}\,,
} 
where $v_{i}$ is the group velocity of each mode at the central frequency $\bar \omega_{i}$. 

The (interaction picture) Hamiltonian governing this three wave mixing process is \cite{quesada2014effects,quesada2017}:
\eq{\label{3wm} 
 	H_I(t) = -\hbar \varepsilon \int d\omega_{1} d \omega_{2} d\omega_{0} e^{i(\omega_{1}+\omega_{2} -\omega_{0}) t}  \Phi\left(\frac{\Delta k(\omega_{0},\omega_{1},\omega_{2}) L}{2} \right ) c_{0}(\omega_{0}) c_{1}^\dagger(\omega_{1}) c_{2}^\dagger(\omega_{2})  +\hc\,,
 }
 where 
 \eq{\label{pm}
 	\Delta k(\omega_{0},\omega_{1},\omega_{2})=k_{0}(\omega_{0})-k_{1}(\omega_{1})-k_{2}(\omega_{2}),
 } 
 and $\Phi$ is the (dimensionless) phase-matching function defined in terms of the (dimensionless) spatial profile  $\chi(z)$ of the nonlinear region extending from $-L/2$ to $L/2$:
 \eq{\label{pmf}
 	\Phi\left(\frac{\Delta k L}{2} \right) =\int \frac{dz}{L} e^{i \Delta k z} \chi(z),
 }
 and the  destruction and creation operators satisfy the usual commutation relations 
 \sseq{\label{ccr}
 	[c_{i}(\omega),c_{j}^\dagger(\omega')]&=\delta_{i,j}\delta(\omega-\omega'),\\
 	[c_{i}(\omega),c_{j}(\omega')]&=[c_{i}^\dagger(\omega),c_{j}^\dagger(\omega')]=0.
 }
If the nonlinear region has a top-hat profile $\chi(z)=1$ then the phase-matching function has the familiar shape $\Phi(x) = \text{sinc}(x)$.\\

\subsection{Spontaneous Parametric Down-Conversion}\label{sec:SPDC}
If the pump  mode ($i=0$) is prepared in a strong coherent state that remains undepleted during the interaction, one can  replace $c_{0}(\omega_{0}) \to \braket{c_{0}(\omega_{0})}=\sqrt{N} \Gamma(\omega_{0})$ where $N$ is the mean number of photons in mode $0$ and $\Gamma$ is proportional to the pump electric field and is normalized according to $\int d\omega |\Gamma(\omega)|^2 = 1$. This   implies that the units of $\Gamma$ are the same as $\omega^{-1/2}$, and the same as the operator $c$.
To first order in perturbation theory one can approximate the state of the downconverted modes generated in SPDC as \cite{grice1997spectral}

\eq{
	\ket{\psi}&=\mathcal{T}\exp\left( \frac{-i}{\hbar} \int_{-\infty}^{\infty} dt H_I(t) \right) \ket{\text{vac}}\\\label{eq:approxspdc}
	&\approx \left(\mathbb{I}+2\pi \sqrt{N} i \varepsilon  \int d\omega_{1} d\omega_{2} \ \Phi\left(\Delta k(\omega_{1}+\omega_{2},\omega_{1},\omega_{2})\frac{L}{2} \right) \Gamma(\omega_{1}+\omega_{2}) c_{1}^\dagger(\omega_{1}) c_{2}^\dagger(\omega_{2} )  \right) \ket{\text{vac}}, 
}

where we used the fact that $\int dt e^{i \Delta t} = 2 \pi \delta(\Delta)$,  $\delta(x)$ is the Dirac distribution, $\mathcal{T}$ is the time-ordering operator, and we assumed that  modes $1$ and $2$ were prepared in the vacuum state $\vac$ before interaction. The approximation in Eq. (\ref{eq:approxspdc}) is only valid in the low-gain limit. In the high-gain regime, the higher-order terms that appear are complicated functionals of the pump and phase-matching function \cite{Branczyk2011b,quesada2014effects,quesada2015time}.

The phase mismatch can be written in simpler terms by using the (simplified) linear dispersion (Eq. \ref{disp})
 \eq{
 	\Delta k(\omega_{1}+\omega_{2},\omega_{1},\omega_{2})\frac{L}{2} =&  
 	\left(\frac{L}{2 v_{1}} -\frac{L}{2 v_{0}} \right) (\omega_{1} -\bar \omega_{1})+\left(\frac{L}{2v_{2}} -\frac{L}{2 v_{0}} \right) (\omega_{2} -\bar \omega_{2}) \nonumber
 	.
 }
Finally, we assume that the pump is prepared in a strong coherent state with carrier frequency $\bar \omega_{0}$ and having some characteristic time scale $\tau$ (the pulse ``duration'') thus one can write 
\eq{\label{betadef}
\Gamma(\omega_0) = \sqrt{\tau}\gamma(\tau(\omega_0 - \bar \omega_{0}))
} 
The prefactors in the last equation guarantee that $\int dy |\gamma(y)|^2=1$, where $\gamma$ is also dimensionless.

Using the characteristic time scale $\tau$ just introduced  we can write the problem in dimensionless form. To this end introduce dimensionless variables 
\sseq{
	\wo &= \tau (\omega_{1}-\bar \omega_{1}), \quad \wt = \tau (\omega_{2}-\bar \omega_{2}),\\
	r &= \frac{L}{\tau}\left(\frac{1}{2 v_{1}} -\frac{1}{2 v_0} \right), \quad
	s = \frac{L}{\tau}\left(\frac{1}{2 v_{2}} -\frac{1}{2 v_0} \right),
}
and then we can write the phase mismatch as
\eq{
	\Delta k(\omega_{1}+\omega_{2},\omega_{1},\omega_{2})\frac{L}{2} = r \wo+ s \wt
	} 
Furthermore we can write the downconverted ket as
\eq{\label{lowestket}
	\ket{\psi}\approx  \ket{\text{vac}}+ \mathcal{N}\ket{\text{II}}
	}
	where
\eq{\label{eq:blahd}
	\mathcal{N}\ket{\text{II}}&\approx 2\pi \sqrt{\frac{N }{\tau}} i \varepsilon\int d x d y \ \Phi\left(  r x+ s y  \right) \gamma(x + y) c_{1}^\dagger( x ) c_{2}^\dagger( y )   \ket{\text{vac}}. 
}

Eq. (\ref{lowestket}) makes explicit that to lowest order in perturbation theory the state of the downconverted modes is mostly vacuum with a small ($\mathcal{N}$) amplitude of a two photon state $\ket{\text{II}}$.
With these simplifications we can finally write the joint spectral amplitude (JSA) of the photons as
\eq{\label{JSA}
	J(\wo,\wt)  = 2 \pi \sqrt{\frac{N}{ \tau}} \varepsilon \Phi (r \wo+s \wt) \gamma(\wo+\wt).
}
We can calculate the squared norm of the JSA as follows
\sseq{
	\mathcal{N}^2&=\int d x d y |J(\wo,\wt)|^2 \\
	&= 4 \pi^2 \frac{N}{ \tau} \varepsilon^2 \int d \wo d\wt | \Phi (r \wo+s \wt) \gamma(\wo+\wt)|^2 \\
	&= 4 \pi^2 \frac{N}{ \tau} \varepsilon^2 \int d u d v \left|\frac{\partial(\wo,\wt)}{\partial (u,v)} \right|  |\gamma(u)|^2 | \Phi(v) |^2.
}
In the last equation we used the change of variables $u=r \wo+s \wt$ , $v=\wo+\wt$ and the Jacobian of the change of variables $\left|\frac{\partial(\wo,\wt)}{\partial (u,v)} \right| = \frac{1}{|r-s|}.$ Remembering that $\gamma$ was already $L^2$ normalized 
\eq{
	\mathcal{N}^2	&= 4 \pi^2 \frac{N}{ \tau} \varepsilon^2 \frac{1}{|r-s|} \underbrace{\int dv|\Phi(v)|^2}_{ \equiv \mathcal{M}^2}=4 \pi^2 \frac{N}{ \tau} \varepsilon^2 \frac{1}{|r-s|} \mathcal{M}^2.
}
As already mentioned, the quantity $\mathcal{N}$ (assumed to be $\ll 1$, otherwise we can not use perturbation theory) gives us the probability of making a pair of photons in modes $1$ and $2$ each time a pump pulse goes through the nonlinear region.
It is convenient to work with normalized quantities thus we write the normalized JSA as
\seq{\label{biph}\eq{
	\Psi(\wo,\wt) &= \frac{J(\wo,\wt)}{\mathcal{N}} = \sqrt{\frac{|r-s|}{\mathcal{M}^2}} \Phi(r \wo+s \wt) \gamma(\wo+\wt) \\
	&= \frac{\sqrt{|r-s|}}{\mathcal{M}} \Phi(r \wo+s \wt) \gamma(\wo+\wt) \\
	&= \sqrt{|r-s|} \phi(r \wo+s \wt) \gamma(\wo+\wt),}
}
which is a properly normalized (biphoton) wavefunction of two quantum systems. In the last equation we introduced the $L^2$ normalized PMF $\phi(x) = \Phi(x)/\mathcal{M}$.
Note that the function defined above becomes zero if $r = s$, if this happens some of the assumptions used to derive this equation are not valid. In particular, one cannot longer ignore group velocity dispersion in Eq. (\ref{disp}).

\subsection{Frequency Conversion}\label{sec:fc}
Going back to Eq. (\ref{3wm}) if one instead prepares mode $2$ in a bright coherent state
\eq{
	c_2(\omega_2) \to \braket{c_2(\omega_2)	} = \sqrt{N} \Gamma(\omega_2),
}
where
\eq{\label{betadefd}
	\Gamma(\omega_2) = \sqrt{\tau}\gamma(\tau(\omega_2 - \bar \omega_{2})).
} 
We can now write the time evolution operator as

\eq{
	\mathcal{U}_{\text{FC}}  &= \mathcal{T} \exp\left(\frac{-i}{\hbar} \int_{-\infty}^{\infty} dt' H_I(t')\right)\\
	&\approx \exp\left( 2\pi \sqrt{N} i \varepsilon  \int d\omega_{0} d\omega_{1} \ \Phi\left(\Delta k(\omega_{0},\omega_{1},\omega_{0}+\omega_{1})\frac{L}{2} \right)  c_{0}(\omega_{0}) c_{1}^\dagger(\omega_{1} ) \Gamma^*(\omega_{0}+\omega_{1}) -\hc \right)
}

where the phase missmatch and phase-matching functions are as defined in Eq. (\ref{pm}) and Eq. (\ref{pmf}). In the last Eq. we introduced a frequency conversion gate $\mathcal{U}_{\text{FC}}$ and assumed time ordering corrections to be negligible \cite{quesada2016high}, this implies that a single application of $\mathcal{U}_{\text{FC}}$ will not approach unit conversion efficiency. However one can concatenate several $\mathcal{U}_{\text{FC}}$ to attain unit conversion efficiency \cite{reddy2014efficient,reddy2013temporal,reddy2017engineering,christensen2015temporal}. As done for SPDC we expand each wavevector as in Eq. (\ref{disp}) and assume the pump mode (mode $2$) has some characteristic time scale $\tau$ and write $\Gamma$ in terms of $\gamma$ as in Eq. (\ref{betadef}). As done for SPDC we introduce dimensionless variables
\eq{
	x=\tau(\omega_0-\bar \omega_0), &\quad 	y=\tau(\omega_1-\bar \omega_1)\\
	r=\frac{L}{\tau}\left(\frac{1}{2 v_0}-\frac{1}{2 v_2} \right),&\quad s=\frac{L}{\tau}\left(\frac{1}{2 v_1}-\frac{1}{2 v_2} \right).
}
Note that in the last expression we take group velocities relative to mode $2$ which is the pump mode for FC and thus $r$ and $s$ are defined differently than for the case of SPDC.
The phase missmatch as now
\eq{
	\Delta k(\omega_0,\omega_1,\omega_0+\omega_1) = r x- s y .
}
We now introduce the joint \emph{conversion} amplitude 
\eq{
	\tilde J(x,y)=2 \pi \sqrt{\frac{N}{\tau}} \varepsilon \Phi(r x- s y ) \gamma^*(x-y).
}
We can construct a normalized transfer function as
\eq{
	\tilde{\Psi}(x,y)= \sqrt{|r-s|} \phi(r x- s y) \gamma^*(x-y)
}
The normalized conversion amplitude $\tilde \Psi$ can be formally obtained from the biphoton wavefunction in Eq. (\ref{biph}) by taking $\gamma \to  \gamma^*$ and $y \to -y$.

\subsection{Mapping to $\chi^{(3)}$ processes }\label{sec:dp}

Even though all the results in this manuscript are written for $\chi^{(2)}$ they carry over to $\chi^{(3)}$ processes where only one pump is used. This is because one can map the $\chi^{(3)}$ Hamiltonian in the undepleted pump approximation to the $\chi^{(2)}$ Hamiltonian used here (see Appendix 3 of Ref. \cite{quesada2014effects}). The difference between the two reduces to the fact that $\Gamma$ which is the pump amplitude in a $\chi^{(2)}$ is the pump amplitude \emph{squared} in a $\chi^{(3)}$ process. Because of this all the results derived in the main text carry over to single pump $\chi^{(3)}$; this is not the case to dual pump $\chi^{(3)}$ schemes \cite{fang2013state,reddy2014efficient,Silverstone2014,Helt2017}.

\section{Review of Harmonic Oscillators}\label{review}

For two harmonic oscillators with raising and lowering operators $\hat a$, $\hat b$ satisfying the canonical commutation relation:
\eq{
[\hat a,\hat a^\dagger]=[\hat b,\hat b^\dagger]=&1, \\
[\hat a,\hat b] = [\hat a^\dagger , \hat b] = [\hat a^\dagger ,\hat b^\dagger ] = [\hat a , \hat b^\dagger]=&0,
} one can define ``position'' and ``momentum'' operators according to
\eq{
	\hat x=\frac{1}{\sqrt{2}} \left(\hat a+\hat a^\dagger\right), &\quad \hat p=\frac{i}{\sqrt{2}} \left(\hat a^\dagger - \hat a\right),\\
	\hat y=\frac{1}{\sqrt{2}} \left(\hat b+\hat b ^\dagger\right), &\quad \hat q=\frac{i}{\sqrt{2}} \left(\hat b^\dagger - \hat b\right).
}

The quadrature operators satisfy the usual commutation relations $[\hat x, \hat p]=[\hat y,\hat q]=i$ and have a continuous spectrum. For example for the $x$ quadrature one has 
\eq{
	\hat x \ket{x} = x \ket{x}, \quad \langle x|x' \rangle =\delta(x-x') \text{ and } \int dx \ket{x} \bra{x} =  \mathbb{\hat I}.
}
For every state of the Harmonic oscillator we associate a wavefunction according to $\ket{\psi} \to \braket{x|\psi}=\psi(x)$.

We now introduce two linear optical operations: the squeezing operation and the beam splitter operation.

\subsection{Squeezing operation}

For mode $a$ a squeezing operation is defined by the unitary operator
\eq{
	\hat S(\mu) = \exp\left(\frac{\ln(\mu)}{2}(\hat a^2 -\hat a^{\dagger2}) \right),
}
where $0<\mu =\exp(r)$, and $r  \in \mathbb{R}$ is the usual squeezing parameter. Notice that in this manuscript we parametrize squeezing by the exponential of the squeezing parameter which corresponds to how much the $x$ axis is squeezed $\mu>1$ or  anti-squeezed $\mu<1$.
This  operator acts in a particularly simple way on the eigenkets of position (see Sec. 8.1.2 of Ref. \cite{kok2010introduction})
\eq{
	\hat S(\mu) \ket{x} &= \sqrt{1/\mu}\ket{ x/\mu},\\
	\bra{x}\hat S(\mu)  &= \sqrt{\mu}\bra{ x \mu}.
}
Also note that $\hat S(\mu)^\dagger = S(1/\mu)$.
With the map just defined it is easy to see that
\eq{
	\bra{x} \hat S(\mu) \ket{\psi} &= \sqrt{\mu} \ \psi(x \mu).
}
For the squeezed (vacuum) state one has
\eq{\label{sqvac}
	\bra{x} \hat S(\mu) \ket{0} = \frac{1}{\sqrt[4]{\pi/ \mu^2}} \exp\left( - \frac{ (\mu x)^2}{2 } \right),
}
where $\ket{0}$ is the vacuum state whose wavefunction is obtained by setting $\mu=1$ in the last equation.

Note that $S(\mu)$ is a unitary operation and does not change the normalization of the wavefunctions
\sseq{
	\bra{\psi} \hat S(\mu)^\dagger \hat S(\mu) \ket{\psi} &= \bra{\psi} \hat S(\mu)^\dagger \int dx \ket{x} \bra{x} \hat S(\mu) \ket{\psi}  \\
	&=
	\int dx \bra{\psi} \hat S(\mu)^\dagger \ket{x} \bra{x} \hat S(\mu) \ket{\psi}\\
	& =\int dx \ \mu |\psi(\mu x) |^2 \\
	&= \int dz |\psi(z)|^2 \\
	&= \int dy \braket{\psi|z} \braket{z|\psi} \\
	&= \braket{\psi|\psi}, 
}
where we made the change of variables $z = \mu x$.

\subsection{Beam splitter operation}

The beam splitter operation  between modes $a$ and $b$ is given by the operator
\eq{
	\uu_{\text{BS}}(\theta) = \exp\left(\theta \left(\hat a^\dagger \hat b - \hat a \hat b^\dagger  \right) \right)
	}
 which acts on the quadrature operators of the two modes according to 
\seq{
	\label{BS}
	\eq{
		\mathcal{\hat U}_{\text{BS}}(\theta)^\dagger & \left(
		\begin{array}{c}
			\hat x   \\
			\hat y
		\end{array}
		\right)
		\mathcal{\hat U}_{\text{BS}}(\theta) = U_{\text{BS}}(\theta) \left(
		\begin{array}{c}
			\hat x   \\
			\hat y
		\end{array}
		\right),\\
		U_{\text{BS}}(\theta)&=\left(
		\begin{array}{cc}
			\cos(\theta) & -\sin(\theta)   \\
			\sin(\theta) & \cos(\theta)   \\
		\end{array}
		\right),\\
		\mathcal{U}_{\text{BS}}(\theta)^\dagger&= \mathcal{U}_{\text{BS}}(-\theta), \quad {U}_{\text{BS}}(\theta)^\dagger={U}_{\text{BS}}(-\theta). \label{dagger}
	}
}
When $\mathcal{\hat U}_{\text{BS}}(\theta)$ is acted  onto a product of position eigenbras or  eigenkets they will transform according to
\sseq{\label{BSaction}
	&\mathcal{\hat U}_{\text{BS}}(\theta) \ket{x} \otimes \ket{y} = \ket{\cos(\theta) x+\sin(\theta)y} \otimes \ket{\cos(\theta) y-\sin(\theta)x}, \\
	&\bra{x} \otimes \bra{y} \mathcal{\hat U}_{\text{BS}}(\theta)= \bra{\cos(\theta) x-\sin(\theta)y} \otimes \bra{\cos(\theta) y+\sin(\theta)x} .
}
Note the difference in sign in the right hand side. This is because of Eq. (\ref{dagger}). Also note that these states do not get entangled after the operation of the BS, this is because the eigenkets $\ket{x}$ are nothing but infinitely squeezed states, $\ket{x} \sim \lim_{\mu \to \infty} \hat S(\mu) \ket{0}$ in Eq. \ref{sqvac}. This is because in our conventions the beam splitter matrix $U_{\text{BS}}$ is real. Because of this the squeezed quadratures $\hat x, \hat y$ are mapped to linear combinations of $\hat x, \hat y$ and are never mixed with the orthogonal quadratures $\hat p, \hat q$.

For a system prepared in a product state we find that its wavefunction becomes 
\sseq{\label{eq:bsss}
\bra{x} \otimes \bra{y} \mathcal{\hat U}_{\text{BS}}(\theta) \ket{\psi} \otimes \ket{\psi'}
	& = \left(\bra{\cos(\theta) x-\sin(\theta)y} \otimes \bra{\cos(\theta) y+\sin(\theta)x} \right)\ket{\psi} \otimes \ket{\psi'}\\
	&=\psi(\cos(\theta) x-\sin(\theta)y) \psi'(\cos(\theta) y+\sin(\theta)x).
}
Given our beam splitter convention the only state that does not get entangled after going through a beam splitter is the state $\left( \hat S(\mu) \otimes \hat S(\mu) \right) \ket{0} \otimes \ket{0} $ where $\ket{0}$ is the ground state.

\section{Mapping the JSA to the state $\ket{\Psi}$}\label{proof22}


\subsection{Spontaneous parametric downconversion}

Imagine that one has such two states $\ket{ \phi}$ and $\ket{ \gamma}$ that are in a local product state and then undergo local squeezing and beam splitter operations 
\eq{\label{afterU}
	\mathcal{\hat U}_{\text{BS}}(\theta) \left( \hat S(\mu) \otimes \hat S(\nu)  \right) \ket{ \phi} \otimes \ket{\gamma} .
}
Note that if the state after the beam splitter is separable the action of local squeezing operations will not change this property. Thus we will consider the wavefunction of the state in Eq. (\ref{afterU}) after applying two more local squeezing operations with parameters $\kappa$ and $\sigma$.
In the position representation the joint wavefunction is
\eq{\label{lt}
	\Psi(x,y)=&\bra{x}  \otimes \bra{y} \left(\hat S(\kappa) \otimes \hat S(\sigma) \right)\mathcal{\hat U}_{\text{BS}}(\theta) \left(\hat S(\mu) \otimes \hat S(\nu) \right)  \ket{\phi} \otimes \ket{ \gamma}
}

Using Eqs. (\ref{sqvac}) and (\ref{eq:bsss}), we have
\eq{\label{feq}
	\Psi(x,y)= \sqrt{\mu \nu\kappa \sigma} \phi\left( \mu (x \kappa  \cos \theta- y \sigma   \sin \theta) \right) \gamma\left( \nu (y \sigma  \cos \theta+ x\kappa  \sin(\theta))  \right) .
}
 
We want this last equation to look like (Eq. (\ref{biphb})):
\eq{\label{eq:temp}
\Psi(\wo,\wt) &= \sqrt{|r-s|} \phi(r \wo +s \wt) \gamma(\wo+\wt)
}
We thus need
\eq{
	\kappa \cos \theta = r,\\
	-\sigma \sin \theta = s,\\
	\sigma \cos \theta = \kappa \sin \theta.
}
These equations can be easily inverted if $r s < 0$
\sseq{\label{sol1}
	\kappa &= \sqrt{r (r -s)},\\
	\sigma &= \sqrt{s (s -r)},\\
	\tan(\theta) &= \sigma/\kappa = \sqrt{-\frac{s}{r}},
}
and $ \sigma \cos \theta = \kappa \sin \theta= \sqrt{-r s}$, $\kappa \sigma = \sqrt{-r s}|r-s|$. Plugging back into  Eq. (\ref{feq}) we find
\eq{\label{jprod}
	\Psi(x,y) =&  \sqrt{ \mu \nu|r-s| \sqrt{-r s}} \phi\left(\mu(r \wo+s \wt)\right) \gamma\left(\nu \sqrt{-r s} (\wo+\wt)\right) .
}
To complete the mapping we pick 
\sseq{\label{sol2}
	\nu &= 1/\sqrt{-r s},\\
	\mu &=1.
}
To obtain precisely Eq.(\ref{eq:temp}).

The derivation just presented is only valid for $r s < 0$.  Note however that as argued at the end of Sec. \ref{separability} this is precisely the regime where one can have any hope of finding separable biphoton wavefunctions thus no generality is lost by assuming $r s < 0$. 
The special case $r s =0$ is dealt with in Sec. \ref{assym}.

\subsection{Frequency conversion}\label{sec:fcmap}

In this section, we show that the  results for SPDC in Sec. \ref{nogo} can be applied to FC. To determine correlations in the joint conversion transfer function
\eq{
		\tilde{\Psi}(x,y)= \sqrt{|r-s|} \phi(r x- s y) \gamma^*(x- y),
	}
we first map it to the ket
\eq{
	\tilde \Psi(x,y)=(\bra{x} \otimes \bra{y}) \ket{\tilde \Psi};~~\ket{  \tilde \Psi} = \int dx dy \   \tilde \Psi(x,y) \ \ket{x} \otimes \ket{y}	 .
	}
We then write 
\eq{
	\ket{\tilde \Psi}=&    \left(\mathbb{\hat I} \otimes \hat P \right) \left( \hat S(\kappa) \otimes \hat S(\sigma) \right)\mathcal{\hat U}_{\text{BS}}(\theta) \left(  \mathbb{\hat I} \otimes \hat S(\nu) \right) \ket{\phi} \otimes \ket{\gamma^*}, \nonumber
	}
where $\hat P = \exp(i \pi \hat b^\dagger \hat b)$ is the parity operator that, when acted on on position eigenket, performs an inversion
\eq{
	\hat P \ket{y} = \ket{-y}, \quad 	\bra{y} \hat P  = \bra{-y}.
	}
The operation necessary to correct for the sign difference between $\Psi(x,y)$ and $\tilde \Psi(x,y)$ is a local one on mode $\hat b$ thus it does \emph{not} change the entanglement properties. Equivalently the entanglement of $\Psi(x,y)$ and $\tilde \Psi(x,y)$ is the same (modulo the substitution of $\gamma$ by $\gamma^*$).

\end{document}